\documentclass[onecolumn,a4paper,11pt]{IEEEtran}
\onecolumn
\IEEEoverridecommandlockouts
\usepackage{amsmath,epsfig}
\usepackage{setspace}
\usepackage{cite}
\usepackage{graphicx}
\usepackage{subfigure}
\usepackage{caption}
\usepackage{amssymb}

\ifCLASSINFOpdf
\else
\fi
\begin{document}

\pagenumbering{arabic}
\setcounter{page}{1}
\renewcommand{\thetable}{\arabic{table}}

\title{Practical Polar Code Construction Using Generalised Generator Matrices}

\author{\IEEEauthorblockN{Berksan Serbetci and Ali E. Pusane}\\
\IEEEauthorblockA{Department of Electrical and Electronics Engineering\\
Bogazici University\\
Istanbul, Turkey\\
E-mail: \{berksan.serbetci, ali.pusane\}@boun.edu.tr}
}

\maketitle

\thispagestyle{plain}
\pagestyle{plain}

\begin{abstract}
Polar coding is a recently proposed coding technique that can provably achieve the channel capacity. The polar code structure, which is based on the original $2\times 2$ generator matrix, polarises the channels, i.e., a portion of the channel capacities approach $1$, while the remaining channel capacities approach $0$. Due to the specific size of this original generator matrix, polar codes can only have code lengths equal to the powers of 2, resulting in inefficiency for codes of practical lengths. In this paper, the performance of finite-length polar codes over the binary erasure channel is analysed. A normalised polarisation distance measure is defined and polar codes from different generator matrices showing different amount of polarisation are compared using this measure. Encoding structures for these generalised polar codes are proposed and polarisation performances in both asymptotical and finite-length cases are investigated for generator matrices of size $3\times 3$ and $4\times 4$. A generalised decoder is also proposed for this generator matrix and its erasure rate is compared with that of the original generator matrix. It is shown that polar codes that have performance similar to the original construction can be constructed and used for a variety of code lengths, not necessarily equal to powers of 2, using generalised generator matrices.

\end{abstract}

\IEEEpeerreviewmaketitle

\section{Introduction}
\label{intro}


Shannon, in his 1948 paper \cite{S48}, calculated the channel capacity -- the ultimate limit of error-free communication over a noisy channel. Prior to 1990s, attempts on designing capacity achieving coding schemes generally failed, since the infeasible code lengths required for good error performance made it impossible to obtain practical coding schemes. The beginning of the 1990s witnessed the emergence of iteratively decodable codes.
Among these codes, Gallager's low-density parity-check (LDPC) code family \cite{G62,G63} was the most successful one that could \textit{approach} the channel capacity via suboptimal decoding algorithms, e.g., a code family whose decoding threshold is $0.0045 dB$ away from the channel capacity was successfully designed in \cite{CFRU01}. However, the required code length is in the order of several millions and, in general, it is observed that as the code's error performance approaches the channel capacity, the associated error floor rises and the code becomes unusable for high signal-to-noise ratios.

Regardless, the LDPC code family was accepted to be the best and may be the only solution of approaching the channel capacity, until Arikan proposed channel polarisation and the associated polar coding technique to \textit{provably achieve} the channel capacity \cite{A08, A09}. In this approach, a channel cluster undergoes a transformation and the capacity of the channels in the newly obtained channel set \textit{polarises} (for sufficiently large number of channels), i.e., individual channel capacities converge either zero or one. If the channel transformation process is considered as encoding, an information sequence can then be sent over the channels with high capacity. Channel polarisation, in this regard, is the first analytically provable channel capacity achieving method and allows us to obtain the polar code family, which has low encoding and decoding complexities. Error performance of polar codes that are created with a similar technique to digital signal processing techniques such as the fast Fourier transform was analysed in \cite{A08b,MT09}. Realisation and design of practical polar codes, error bounds, and exponents were studied in \cite{KSU09,KSU10,CNL13}. 

In this paper, polarisation performance of various polar code generator matrices is analysed. Our main focus is on the finite-length polarisation behaviour of polarised channels generated by generator matrices of different sizes. These generator matrices are evaluated via both asymptotical polarisation rate exponents and finite-length polarisation measures. As comparing different generator matrices via histogram plots is not an easy task, a finite-length normalised polarisation distance measure is proposed in order to provide distance plots and compare polarisation levels. The generator matrix showing the best polarisation performance for any given code rate can be determined using this measure. Upper bounds on block error probability for the generator matrices with different polarisation levels are given and compared with Arikan's original generator matrix' bound. Moreover, the recursive likelihood ratio equations for a specific $4\times 4$ generator matrix are computed and a decoding algorithm that utilises these equations is implemented. It is shown that using larger generator matrices allows obtaining polar codes with flexible code lengths while maintaining an error performance comparable to that of the original construction. This is an efficient way to obtain polar codes of flexible code lengths and is comparable to the contribution of \cite{CNL13}. In \cite{CNL13}, the authors propose the use of puncturing together with codes with length equal to the next power of 2 to obtain the desired code length. This approach can be used in conjunction with the proposed approach to obtain codes of desired length with minimal use of puncturing by starting with a slightly larger block length (not necessarily the next power of 2) and puncturing the unnecessary symbols/channels. 

The paper is organised as follows: Section II covers the preliminary concepts of polar coding and reviews the standard code construction method. The main contribution of the paper is presented in Section III, where we consider the case of generalised generator matrices of larger sizes and define a normalised polarisation distance measure to quantify the amount of polarisation for finite-length codes. A generalised decoder is also proposed in this section. Section IV concludes the paper.

\section{POLAR CODING}
\label{sec:polarcoding}
For any given binary-input symmetric channel, the channel capacity, $I(W)$, is an easy-to-compute value and satisfies the inequality $0 \leq I(W) \leq 1$.  Based on Shannon's channel capacity theorem, the code rate of a communication channel $0\leq R \leq 1$ also fits into this inequality as $0 \leq R \leq I(W) \leq 1$ for error-free communication. If the two extreme values are considered, i.e., $I(W) = 0$ and $I(W)=1$, it is seen that the code design problem can be simplified. For the $I(W)=0$ case, the code rate satisfies $R=0$ and error-free communication is impossible regardless of the coding scheme employed, hence, no information can be transmitted across the channel. For $I(W)=1$, the code rate can be chosen as $R=1$ and even uncoded bits can be transmitted across the channel without errors. Arikan's {\it channel polarisation} and the associated {\it polar coding} technique exploit these two extreme cases in Shannon's theorem.

The polar coding scheme proposed in \cite{A09} takes $N$ binary-input symmetric channels of capacity $I(W)=\delta$, transforms them (via a recursive structure) to another set of channels with capacity values approaching either $I(W)=0$ or $I(W)=1$. As $N \rightarrow \infty$, $N\delta$ channel capacity values approach $I(W)=1$, whereas the remaining $N(1-\delta)$ channel capacities approach $I(W)=0$. These channels are said to polarise: bad channels do not allow reliable transmission of any data, and good channels allow reliable transmission of uncoded data. Also the total channel capacity of the two sets after polarisation is preserved at $N\delta$. Among the newly created channels, uncoded information symbols can be transmitted over the good channels (with channel capacities approaching $I(W)=1$), whereas bad channels (with channel capacities approaching $I(W)=0$) remain unused. The channel cluster transformation and distribution of information sequence to channels with high capacity are together called polar coding.

\begin{figure}[htbp]
\centering
\includegraphics[width=0.32\columnwidth]{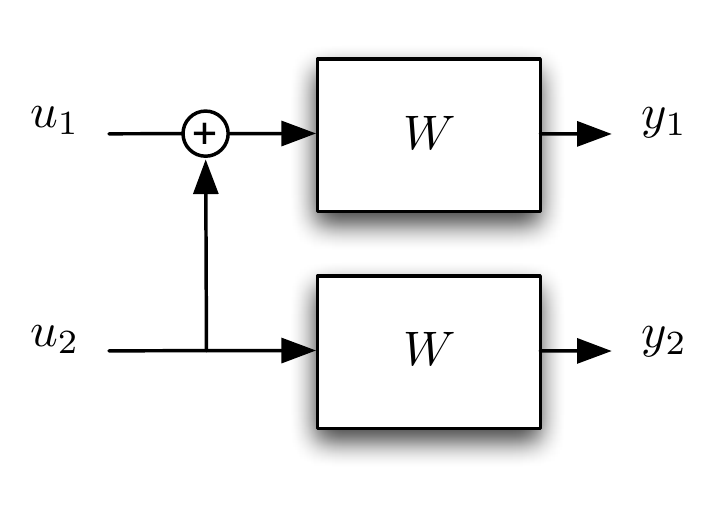}
\caption{Combination of two $W$ channels using $G_2$: the $W_2$ channel.}
\label{fig:w2} 
\end{figure}

The first level of the recursion proposed in \cite{A09} considers the original generator matrix of size $2\times2$. This level corresponds to combining two independent copies of a single channel in order to create $W_2$ using the generator matrix
\[G_{2} = \begin{bmatrix}1 & 0\\ 1 & 1\end{bmatrix}.\] 
The corresponding channel input configuration is shown in Figure \ref{fig:w2}.\\

\begin{figure}[htbp]
\begin{center}
\includegraphics[width=0.5\columnwidth]{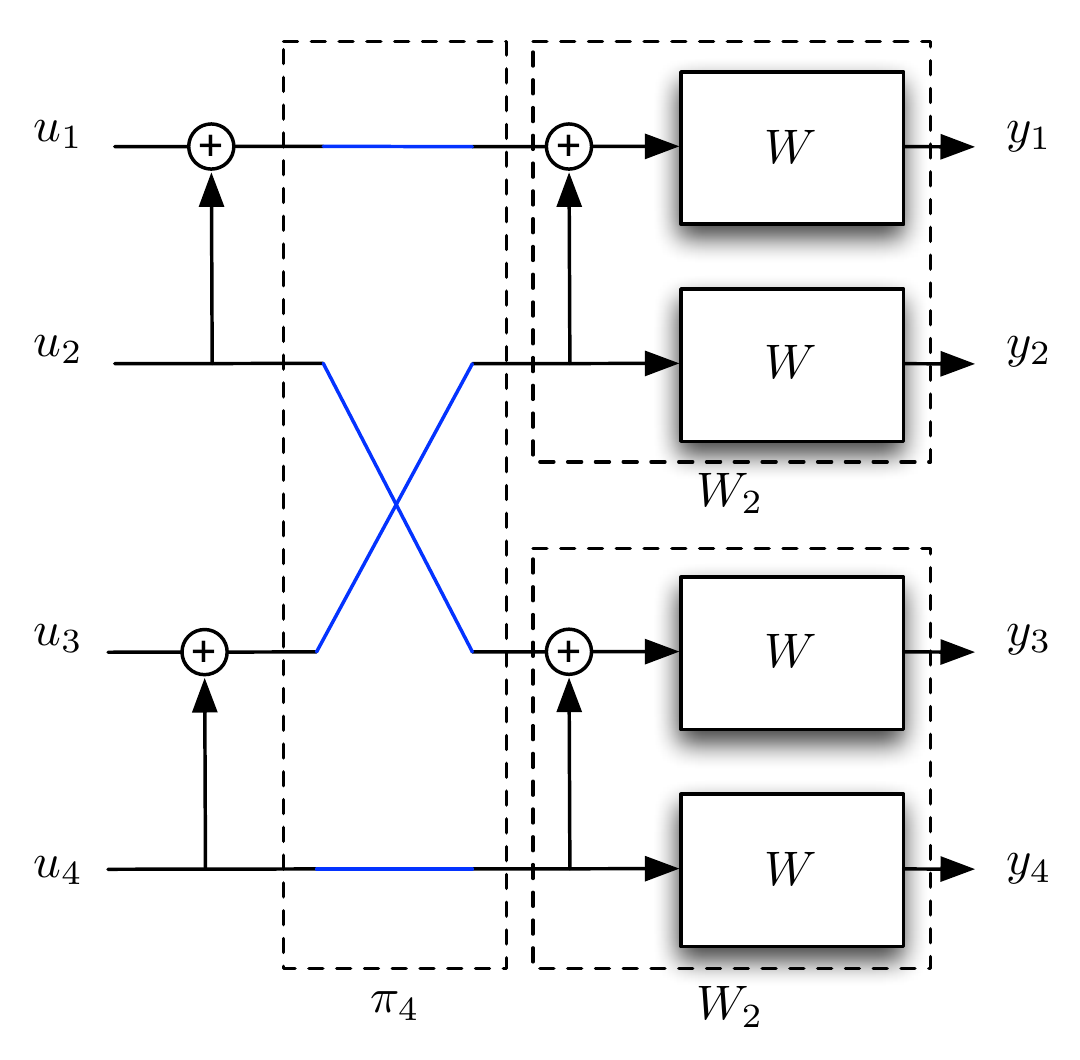}
\end{center}
\caption{The $W_4$ channel.}
\label{fig:w4}
\end{figure}

The next level of recursion consists of two independent copies of $W_2$ combined together to create the channel $W_4: \mathcal{X}^4 \rightarrow \mathcal{Y}^4$ with transition probabilities $W_4(y_1^4|u_1^4)=W_2(y_1^2|u_1\oplus u_2, u_3\oplus u_4)W_2(y_3^4|u_2,u_4)$, where $y_k^l$, $k \leq l$, denotes the vector $(y_k,y_{k+1}, \ldots, y_l)$. The corresponding channel input configuration is shown in Figure \ref{fig:w4}. In this figure, $\pi_4$ is the reverse shuffle operator the maps an input $(s_1,s_2,s_3,s_4)$ to an output $(s_1,s_3,s_2,s_4)$. Its role is to map channels of equal capacity together to the input of the next channel polarization block. In general, the reverse shuffle operator of size $N$, $\pi_N$, maps an input $(s_1,s_2,\ldots, s_N)$ to an output $(s_1,s_3,\ldots, s_{N-1},s_2,s_4,\ldots,s_N)$. In order to obtain larger structures, the generator matrix $G_{2}$ is recursively transformed into larger matrices with a length of desired code sequence using Kronecker products and permutation operators. Specifically, for a desired code length of $N = 2^n$ bits, where $n$ is the number of recursions, the desired matrix is of size $N\times N$ and is obtained by applying the transformation ${G_{2}}^{\otimes n}$ and permutation operation afterwards, where $^{\otimes n}$ denotes the $n^{th}$ degree Kronecker product. This results in a family of polar codes with code lengths of powers of 2.

\begin{figure}
\centering
\includegraphics[width=0.6\columnwidth]{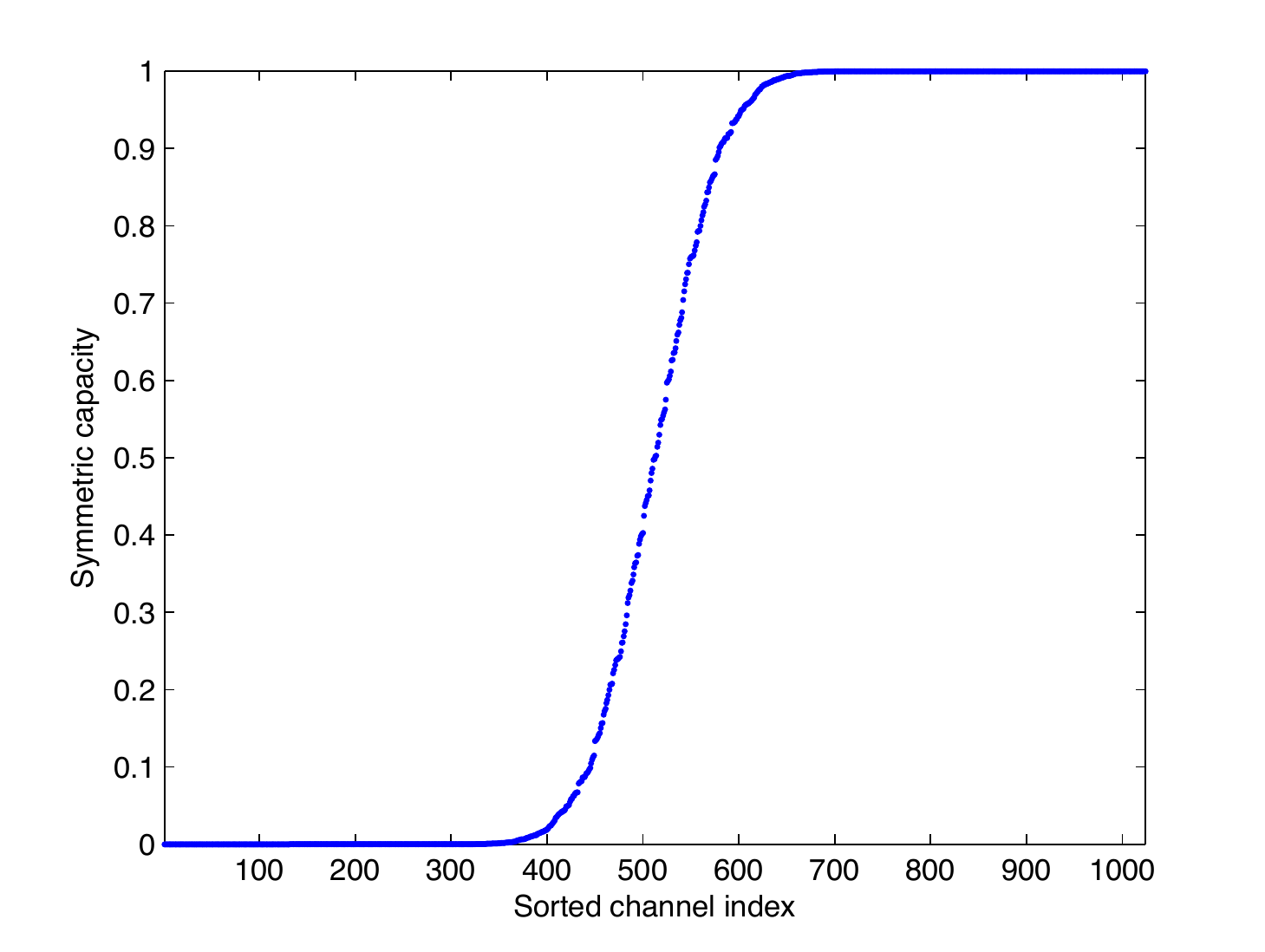}\\
\caption{Polarisation for the generator matrix $G_2$ with $\epsilon = 0.5$.}
\label{fig:2x2} 
\end{figure}

For $N = 2^{10} = 1024$ binary erasure channels (BECs) with channel capacity $I(W)=0.5$, the sorted symmetric capacities obtained using the recursive relations given in \cite{A09} are plotted in Figure \ref{fig:2x2}, where among $N = 1024$ constructed channels, nearly 400 of them have $I(W) \approx 0$, nearly 400 of them have $I(W) \approx 1$, and the rest of the channels have capacities in some mid-values. In the asymptotical case, the ratio of these intermediate channels is expected to go to zero (complete polarisation).

\section{GENERALIZED GENERATOR MATRICES}
\label{sec:genmat}

In \cite{A09}, Arikan states that polarisation is a general phenomenon and does not require a special generator matrix ${G_{2}}^{\otimes n}$ transformation. Korada \textit{et al.}, in \cite{KSU10}, worked on the transformations of generator matrices of size $l\times l$ (for $l\geq 3$) and obtained necessary and sufficient conditions for polarisation. Arikan and Telatar, in \cite{AT09}, showed that, when using the polar coding technique and successive cancellation decoding for the original generator matrix $G_{2}$, the block error probability is $O(2^{-2^{n\beta}})$, where $\beta<\frac{1}{2}$ and $N = 2^n$ is the code length. Therefore, the error exponent of $G_{2}$ is said to be $\frac{1}{2}$. In \cite{KSU09}, it is proved that this exponent can be improved as the generator matrix size increases, potentially approaching 1. This exponent can even approach $1$ when larger generator matrices are used. This result is the main motivation to study larger generator matrices in order to obtain reliable communication using the polar coding technique with low block error probability. In addition to this, by the use of larger generator matrices, we can consider polarisations that are not necessarily spanned by ${G_{2}}^{\otimes n}$.

\subsection{Bhattacharyya Parameters}
\label{subsec:bhat}

In \cite{A09}, Bhattacharyya parameters are employed to obtain the channel capacity values, since these parameters are easily computable. For the original generator matrix $G_2$, Bhattacharyya parameters are computed and recursive channel capacity equations are obtained in order to demonstrate polarisation. Computing these values are a bit more complicated for larger generator matrices and obtaining similar recursive relations is not always easy. The equations to be used while computing the channel transition probabilities at the channel splitting phase are given by
\begin{equation}
W_N^{(i)}(y_1^N,u_1^{i-1}|u_i)=\sum_{{u_{i+1}^N}\in\mathcal{X}^{N-i}}\frac{1}{2^{N-1}}W_N(y_1^N|u_1^N), \label{main_chan}
\end{equation}
where $W_N(y_1^N|u_1^N)$ is the constructed joint vector channel and $W_N^{(i)}(y_1^N,u_1^{i-1}|u_i)$, $i=1,\ldots,n$, is the $i$-th binary-input coordinate channel seen by the $i$-th input symbol given that all observations $y_1^N$ and previous information symbols $u_1^{i-1}$ are known. The subscript $N$ denotes that there are a total of $N$ such formed channels.

The main measure of polarization, the Bhattacharyya parameters, can then be expressed in terms of the channel transition probabilities as
\begin{equation}
Z(W_N^{(i)})=\sum_{y_1^N\in\mathcal{Y}^N}\sum_{u_1^{i-1}\in\mathcal{X}^{i-1}}\sqrt{\prod\limits_{u_i=0,1} W_N^{(i)}(y_1^N,u_1^{i-1}|u_i)}, \label{main_bhat}
\end{equation}
where $\mathcal{X}$ and $\mathcal{Y}$ denote the input and output symbol alphabets, respectively. In this paper, the binary transmission case is considered, i.e., $\mathcal{X}=\mathcal{Y}=\{0,1\}$.

For the example generator matrix of
\[
G_{3} = \begin{bmatrix}1 & 0 & 0\\ 1 & 1 & 0 \\ 0 & 1 & 1\end{bmatrix},
\]
the recursive combination of three independent copies of a single channel has the transition probability given by
\begin{equation}
W_3(y_1^3 | u_1^3) = W(y_1 | u_1 \oplus u_2)W(y_2 | u_2 \oplus u_3)W(y_3 | u_3). \label{W_comb}
\end{equation}

Placing \eqref{W_comb} in \eqref{main_chan}, using \eqref{main_bhat}, and taking into account that  all  channels are binary erasure channels with transition probabilities $W(0|0) = W(1|1) = 1-\epsilon$, $W(0|1) = W(1|0) = 0$, and $W(e|0) = W(e|1) = \epsilon$ ($e$ denotes an erasure), the corresponding Bhattacharyya parameters can be written in terms of $\epsilon$ as
\begin{equation}
Z(W_3^{(1)})=\epsilon^3-3\epsilon^2+3\epsilon,
\end{equation}
\begin{equation}
Z(W_3^{(2)})=-\epsilon^3+2\epsilon^2,
\end{equation}
\begin{equation}
Z(W_3^{(3)})=\epsilon^2.
\end{equation}

For $i=1, \ldots, N$, using (6) in \cite{A09}, calculating the channel capacity recursively is possible, and since $G_3$ is lower-triangular \cite{KSU10}, the equations can be written as
\begin{equation}
I(W_N^{(3i-2)})=I(W_{N/3}^{(i)})^2, 
\end{equation}
\begin{equation}
I(W_N^{(3i-1)})=-I(W_{N/3}^{(i)})^3+ 2I(W_{N/3}^{(i)})^2,
\end{equation}
\begin{equation}
I(W_N^{(3i)})=I(W_{N/3}^{(i)})^3-3I(W_{N/3}^{(i)})^2+3 I(W_{N/3}^{(i)}).
\end{equation}

\begin{figure}
\centering
\includegraphics[width=\columnwidth]{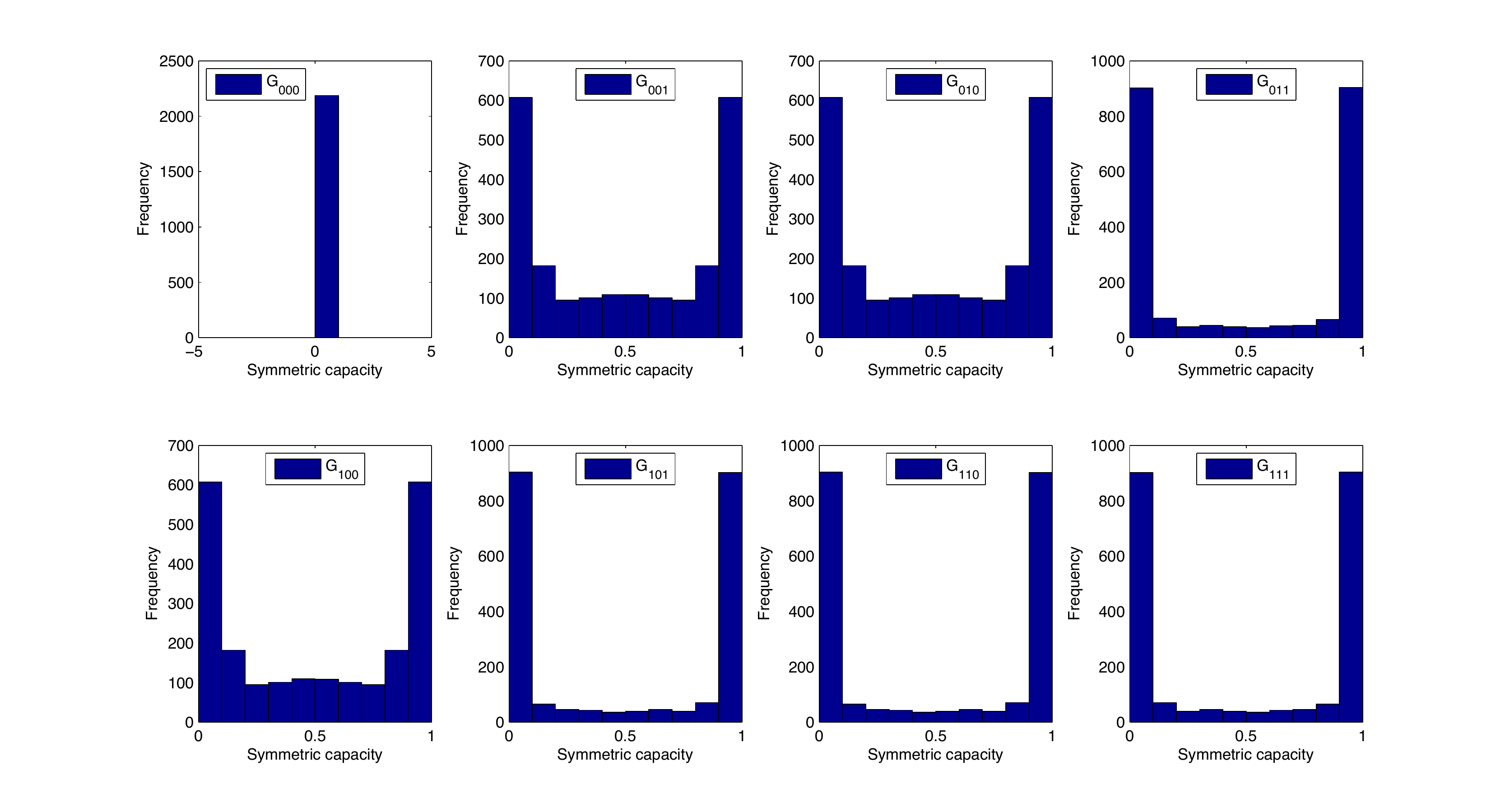}\\
\caption{Polarisation histograms for all lower-triangular $3\times3$ generator matrices with $\epsilon = 0.5$.}
\label{fig:3x3hist}
\end{figure}

Even though computing these values analytically for larger generator matrices is very hard, Bhattacharyya parameters can be numerically calculated. Polarisation histograms corresponding to all possible lower-triangular $3\times3$ generator matrices, denoted by
\[
 G_{abc} = \begin{bmatrix}1 & 0 & 0\\ a & 1 & 0 \\ b & c & 1\end{bmatrix},
\] 
and $N=3^7=2187$ channels over binary erasure channels with $\epsilon =0.5$ are shown in Figure \ref{fig:3x3hist}.

The generator matrices demonstrating the best polarisation performance are $G_{011}$, $G_{101}$, $G_{110}$, and $G_{111}$. Among the remaining matrices, $G_{001}$, $G_{010}$, and $G_{100}$ demonstrate a weaker polarisation, and $G_{000}$ does not show any polarisation. This last observation for $G_{000}$ is expected as this generator matrix corresponds to using three channels independently without any processing of the channel inputs ($G_{000}$ is the identity matrix of size $3\times3$).

\begin{figure}
\centering
\includegraphics[width=0.6\columnwidth]{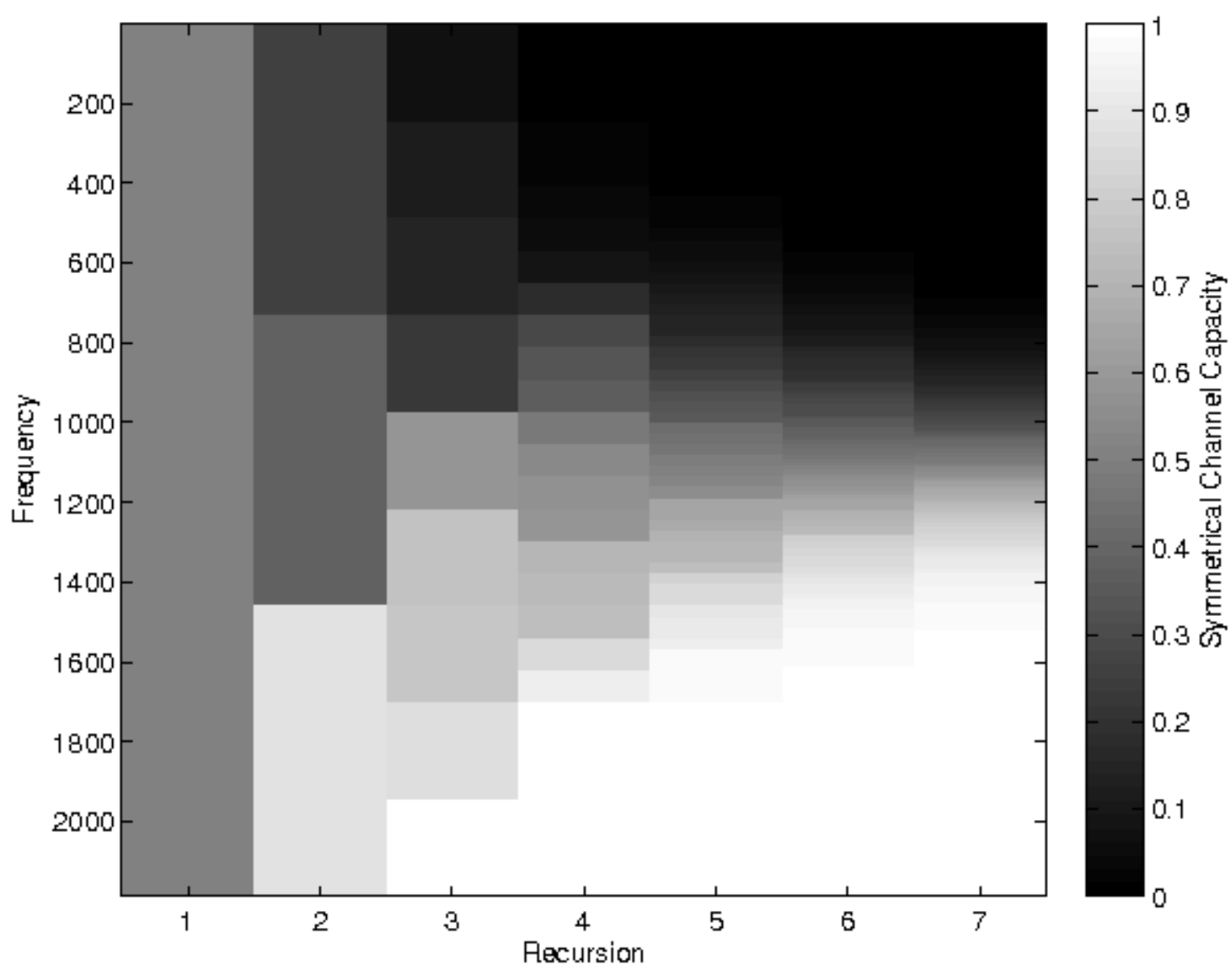}\\
\caption{Evolution of the polarisation tree for $G_{101}$ with $\epsilon = 0.5$.}
\label{fig:evo}
\end{figure}

The polarisation histogram evolution of $G_{101}$ for up to $7$ iterations is also plotted in Figure \ref{fig:evo}. It can be seen that, while the entire histogram consists of channels with erasure values equal to $\epsilon = 0.5$ for the first recursion (the gray colour), the subsequent iterations result in channels with colours polarising towards white ($\epsilon = 1$) and black ($\epsilon = 0$).

\subsection{Polarisation Rate Exponents}
\label{subsec:polrateexp}

It has been shown in \cite{KSU10} that partial distances (Hamming distances between generator matrix rows) can be used as a polarisation performance measure. Specifically, the authors consider the maximisation of the polarisation rate exponent for large generator matrices using 
\begin{equation}
   E_l = \max\limits_{G\in \{0,1\}^{l\times l}} E(G),
\end{equation} 
where $E(G)$ is the polarisation rate exponent for generator matrix $G$ calculated using the partial distances $D_i$ as
\begin{equation}\label{eq:polpar}
  E(G) = \frac{1}{l}\sum_{i=1}^l{log_l{D_i}}.
\end{equation}

Substituting the partial distances $D_1 = 1$, $D_2 = 2$, and $D_3 = 2$ for $G_{101}$ in (\ref{eq:polpar}) yields $E(G_{101}) =\frac{2}{3}log_3(2) = 0.4206$. The polarisation rate exponents for all possible lower-triangular $3\times3$ matrices are similarly obtained using (\ref{eq:polpar}) and presented in Table 1.

\begin{table}
	\caption{Polarisation rate exponents.}
\begin{center}
\begin{tabular}{c r c r}
Matrix & $E_G$ & Matrix & $E_G$\\
\hline
$E(G_{000})$ & $0.000$ & $E(G_{100})$ & $0.210$\\
$E(G_{001})$ & $0.210$ & $E(G_{101})$ & $0.421$\\
$E(G_{010})$ & $0.210$ & $E(G_{110})$ & $0.421$\\
$E(G_{011})$ & $0.333$ & $E(G_{111})$ & $0.333$
\end{tabular}
\end{center}
\end{table}

This result shows that the generator matrices showing the same amount of polarisation based on histogram plot results may have different polarisation rate exponents. Therefore, more than one parameter should be taken into account for the performance evaluation and comparison of finite-length polar coded systems. In order to create the most efficient polar code of a finite length, calculating the partial distances is not always enough. 

\subsection{Normalised Polarisation Distance Measures}
\label{subsec:normpolme}

\begin{figure}[htbp]
\begin{center}
\includegraphics[width=0.7\columnwidth]{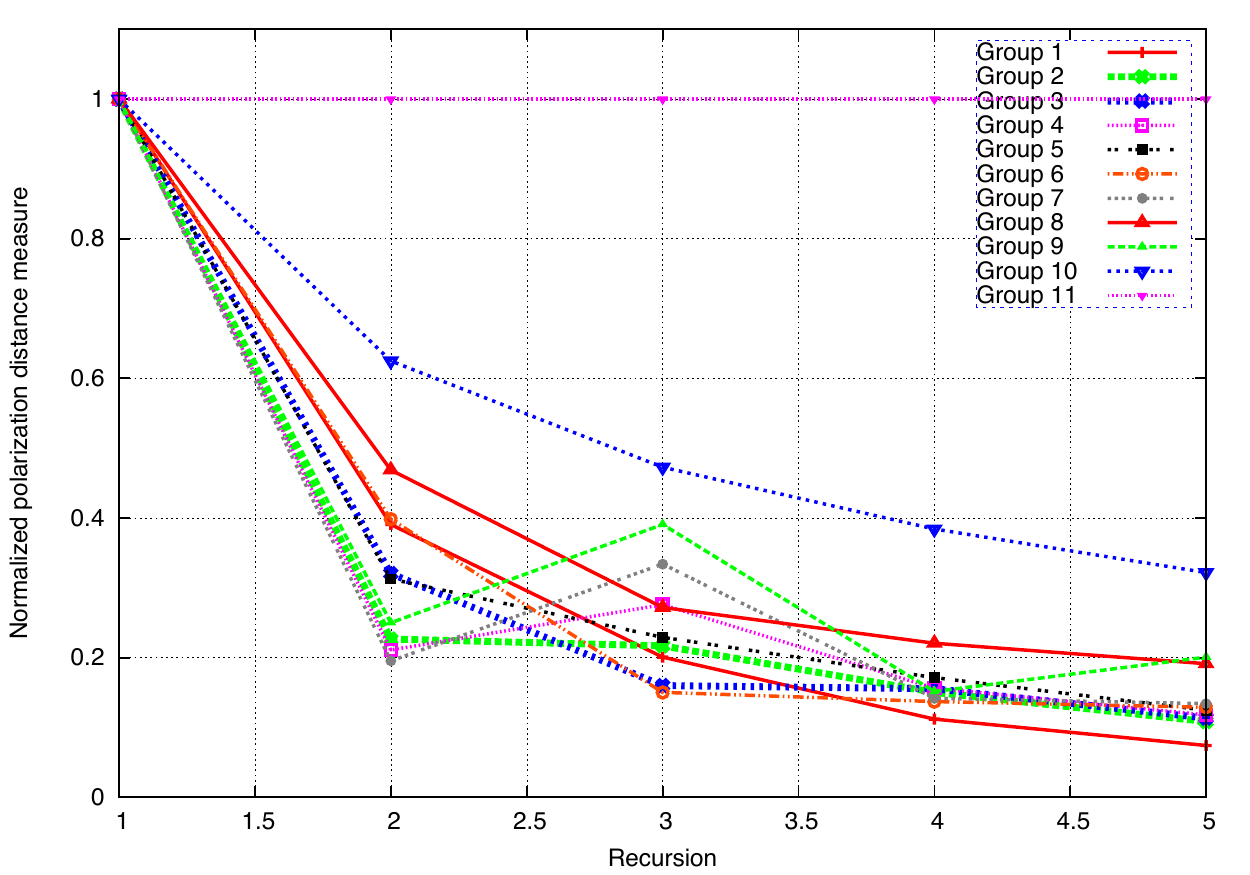}
\end{center}
\caption{Normalised polarisation distance measures for all $4\times 4$ generator matrices with $\epsilon = 0.5$.}
\vskip\baselineskip 
\label{fig:4x4norpoldis}
\end{figure}

Plotting the polarisation histograms and observing the Bhattacharyya parameter distributions for any generator matrix is a useful tool for visually investigating the polarisation behaviour, however, it is not quite helpful for comparing the polarisation of different generator matrices of the same size. In order to come up with a solution to this, in this section, we propose a normalised polarisation distance measure $d_p^{\epsilon_0}(\bar{\epsilon},N)$ for an initial channel erasure probability $\epsilon_0$ and a vector $\bar{\epsilon}=(\epsilon_1,\epsilon_2,\ldots,\epsilon_N)$ consisting of $N$ polarised channel erasure probabilities using the relation
\begin{equation}
d_p^{\epsilon_0}(\bar{\epsilon},N) = \frac{1}{N\epsilon_0^2}\sum_{i=1}^N{\min(|{\epsilon_i}|,|{1-{\epsilon_i}}|)^2}. \label{eq:mindis}
\end{equation}
The normalisation guarantees that the measure satisfies
\[
0\leq d_p^{\epsilon_0}(\bar{\epsilon},N) \leq 1
\]
for all values of $N$, $\epsilon_0$, and $\bar{\epsilon}$. Therefore, the generator matrices showing no polarisation have $d_p^{\epsilon_0}(\bar{\epsilon},N) = 1$ and complete polarisation will result in $d_p^{\epsilon_0}(\bar{\epsilon},N) = 0$. As the normalised polarisation distance measure goes to $0$, the corresponding generator matrices show better polarisation.\\
\begin{figure}[htbp]
\begin{center}
\includegraphics[width=0.7\columnwidth]{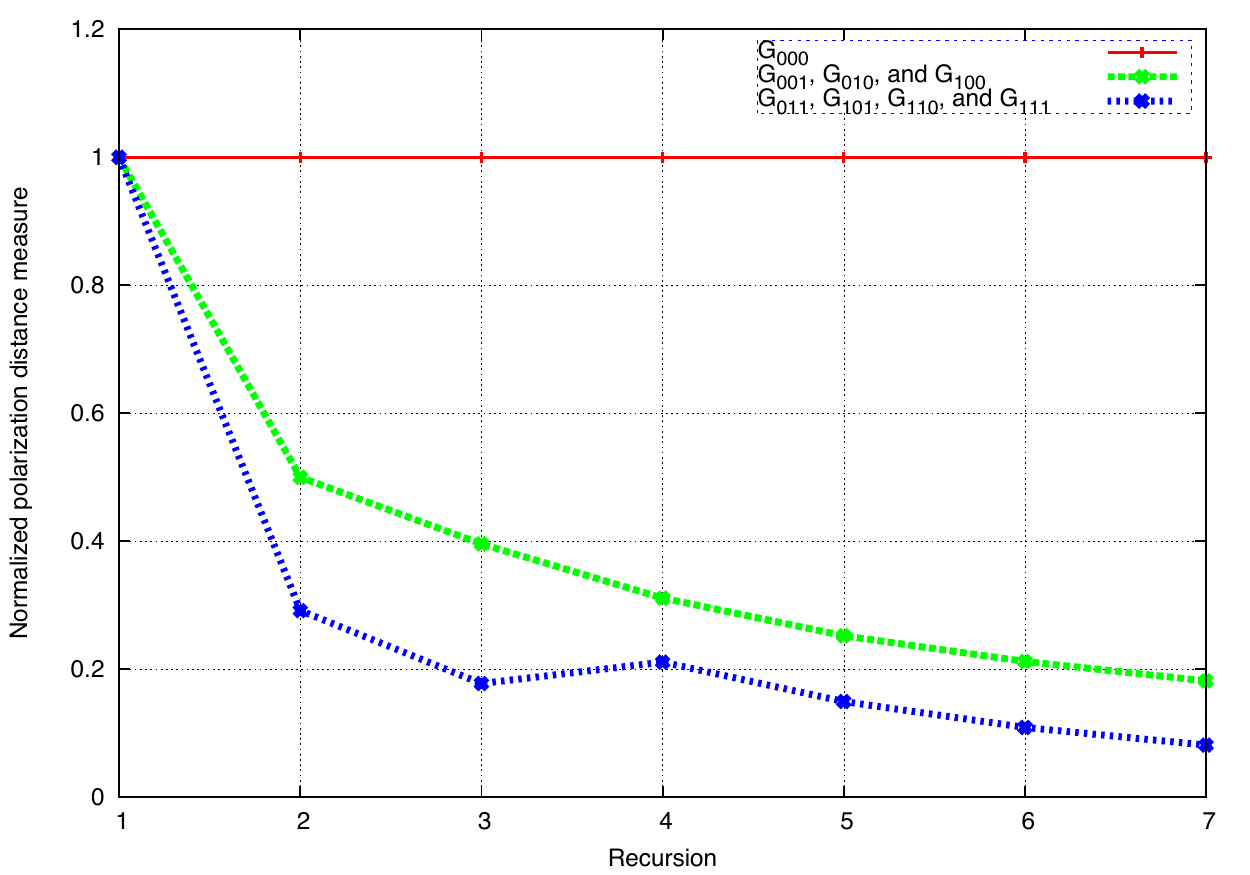}
\end{center}
\caption{Normalised polarisation distance measures for all lower-triangular $3\times 3$ generator matrices with $\epsilon = 0.5$.}
\vskip\baselineskip 
\label{fig:norpoldis}
\end{figure}
Using \eqref{eq:mindis}, the normalised polarisation distance measures of all lower-triangular $3\times3$ generator matrices are plotted in Figure \ref{fig:norpoldis}.
The measure curves corresponding to generator matrices yielding better polarisation fall more rapidly than others as the number of recursions increase (the code length increases). At the extreme case of no polarisation ($G_{000}$), the measure remains constant at the normalised $1$ value, which is the worst polarisation scenario.\\
Again using \eqref{eq:mindis}, the normalised polarisation distance measures for all $4\times 4$ generator matrices are plotted in Figure \ref{fig:4x4norpoldis}. In this case, we observe a grouping behaviour among $2^{16} = 65,536$ generator matrices of size $4\times 4$. There are only $11$ groups of matrices showing different amount of polarisation. Hence, there are $11$ different normalised polarisation distance measure curves.
Only $18,624$ matrices show polarisation among $65,536$ possible $4\times 4$ generator matrices and only $192$ matrices are in Group $1$, whose members show the best polarisation.

\subsection{Upper Bound on Block Error Probability}
\label{subsec:uppbnd}

\begin{figure}[htbp]
\begin{center}
\includegraphics[width=0.8\columnwidth]{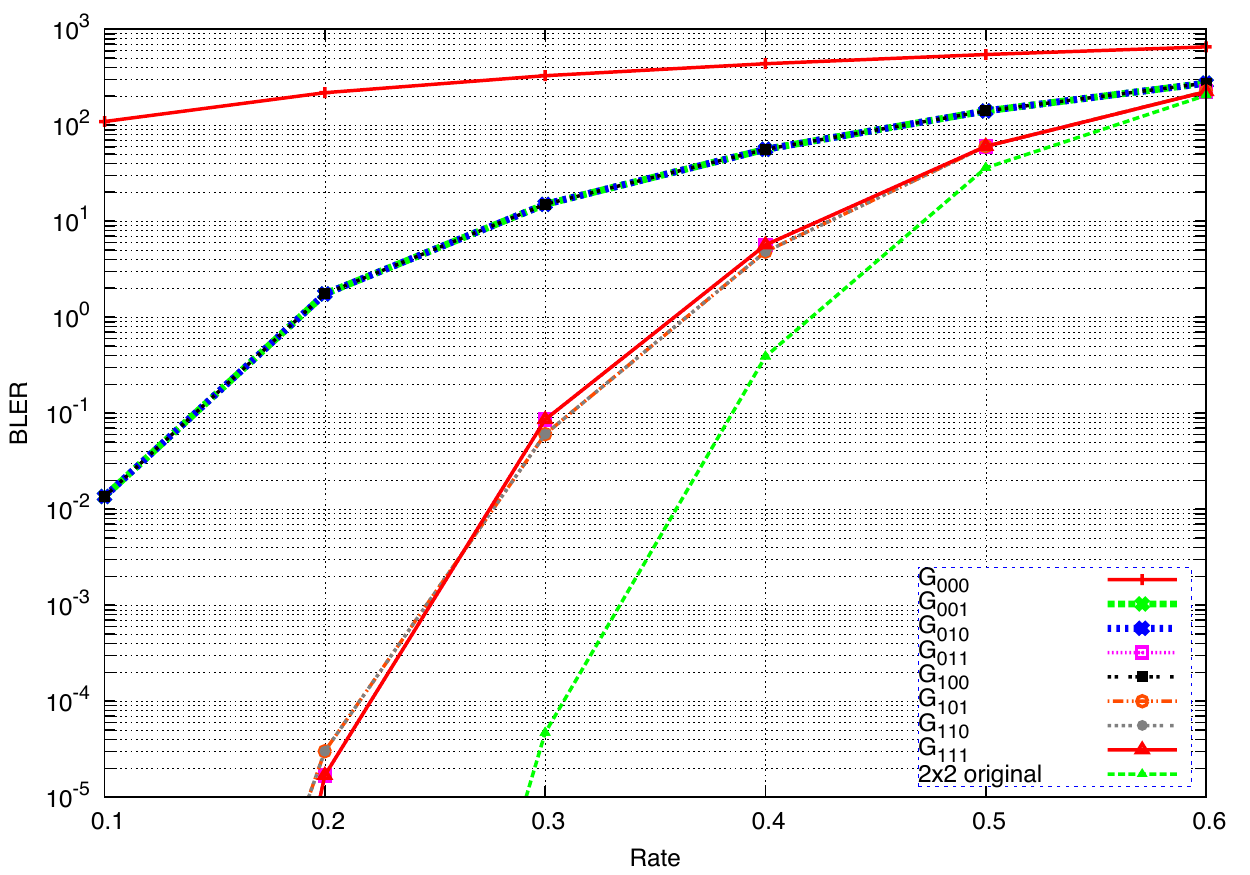}
\end{center}
\caption{Upper bound on block error probability for all $3\times 3$ lower-triangular matrices with $\epsilon=0.5$.}
\vskip\baselineskip 
\label{fig:3x3erruppbnd}
\end{figure}

\begin{figure}[htbp]
\begin{center}
\includegraphics[width=0.8\columnwidth]{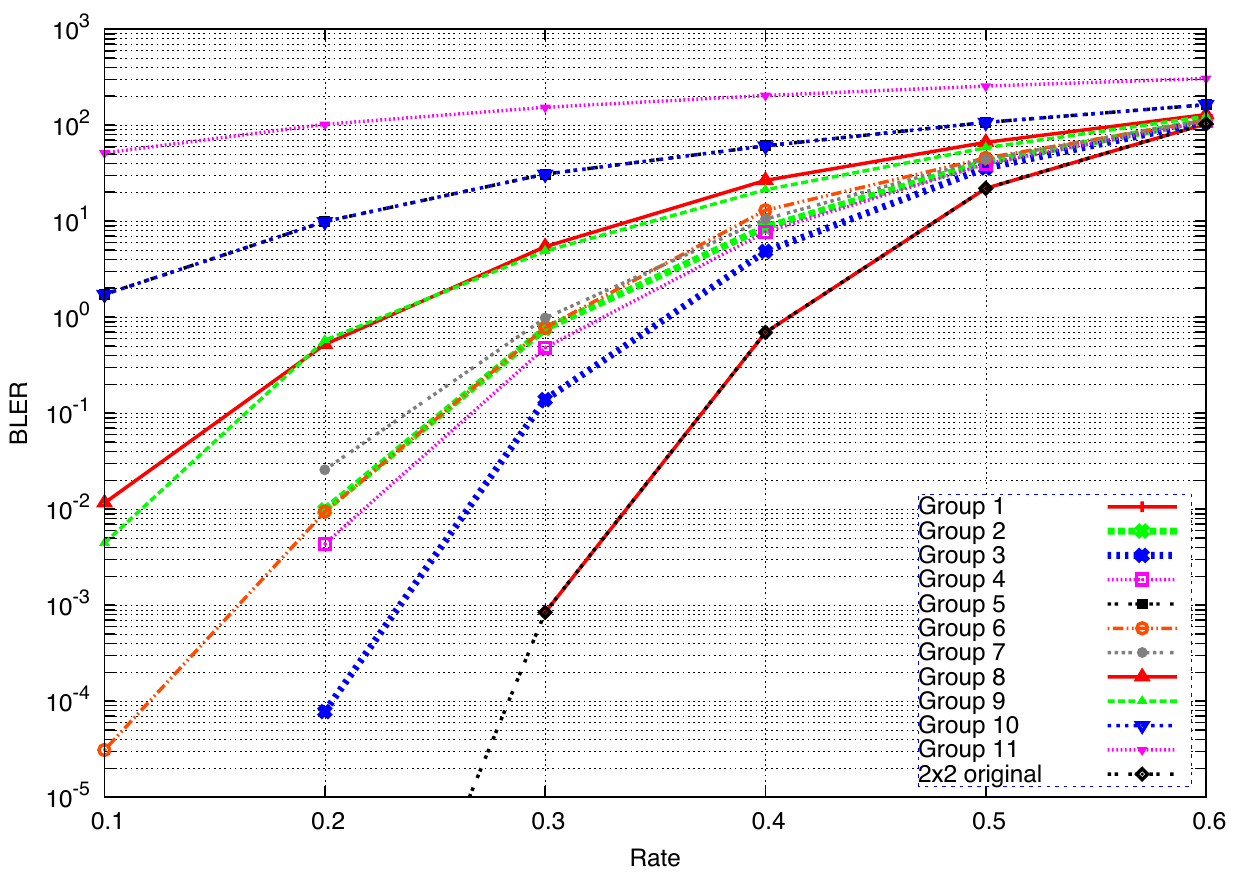}
\end{center}
\caption{Upper bound on block error probability for all $4\times 4$ matrices with $\epsilon=0.5$.}
\vskip\baselineskip 
\label{fig:4x4erruppbnd}
\end{figure}

Using (13) in \cite{A09}, upper bounds on block error probability (BLER) for all $3\times 3$ lower-triangular matrices with $\epsilon=0.5$ are obtained and shown in Figure \ref{fig:3x3erruppbnd}.
The original upper bound for Arikan's $2\times 2$ code is computed at a code length of $N=2^{11}=2048$, whereas the bounds for $3\times 3$ generator matrices are computed at a code length of $N=3^7=2187$. Therefore, an exact comparison is not quite possible between the $2\times2$ and $3\times3$ generator matrices, but it is viable to say that the matrices showing better polarisation also have a more strict bound on block error probability. Moreover, the generator matrices showing the same characteristics on polarisation histograms also have the same upper bounds on block error.\\
Upper bounds on block error probability for all $4\times 4$ generator matrices with $\epsilon=0.5$ are shown in Figure \ref{fig:4x4erruppbnd}.
The original upper bound for Arikan's $2\times 2$ code is computed at a code length of $N=2^{10}=1024$, and the number of recursions for $4\times 4$ generator matrices is chosen as 5 to obtain a code length of $N=4^5=1024$. Again, Group $1$ is the group showing the best polarisation performance, and its upper bound on block error probability curve overlaps with Arikan's curve. Therefore, it is possible to say that choosing a generator matrix from Group $1$ and applying polar code structure is expected to show good block error performance. This is the main reason why a generator matrix from Group $1$ is chosen for the investigation of decoding performance and the implementation of a decoder in the next section.

\subsection{Decoder}
\label{subsec:dec}
For any generator matrix, the likelihood ratio equations need to be computed in order to use the successive cancellation decoder. However, direct computation of $W_N^{(i)}$ values at large block-lengths is impossible. For instance, for $i = 1$, while computing $W_N^{(1)}$ using (\ref{main_chan}), the summation starts from $u_2$ and ends at $u_N$. That is to say, all possible binary combinations should be computed for $2^{N-i}$ values (namely, for the first step, from $u_2$ to $u_N$, $2^{N-1}$ values should be computed and so on). Therefore, a recursive structure is necessary for each generator matrix to avoid this cumbersome task.\\
In order to demonstrate the derivation of this recursive structure, a generator matrix from the best group according to normalised polarisation distance measure is chosen as
\begin{center}
$G_e = \begin{bmatrix}
1 & 0 & 0 & 0\\
1 & 0 & 0 & 1\\
0 & 1 & 0 & 1\\
1 & 1 & 1  &1
\end{bmatrix}$.
\end{center}
The transition  probability of the combined channel is then given as
\begin{equation}
W_4(y_1^4|u_1^4)=W(y_1|u_1\oplus u_2\oplus u_4)W(y_2|u_3\oplus u_4)W(y_3|u_4)W(y_4|u_2\oplus u_3\oplus u_4).
\label{eq:W4g1}
\end{equation}
After obtaining the combined channel transition probability for the specified matrix, channels must be split back into a set of $4$ binary-input coordinate channels $W_4^{(i)}$, $i=1,2,3,4$, whose transition probabilities can be computed as
\begin{align}
W_4^{(1)}(y_1^4|u_1)&=\sum_{u'_2,u'_3,u'_4}\frac{1}{8}W(y_1|u_1\oplus u_2'\oplus u_4')W(y_2|u_3'\oplus u_4')W(y_3|u_4')W(y_4|u_2'\oplus u_3'\oplus u_4'),\label{eq:W41}\\
W_4^{(2)}(y_1^4,u_1|u_2)&=\sum_{u'_3,u'_4}\frac{1}{8}W(y_1|u_1\oplus u_2\oplus u_4')W(y_2|u_3'\oplus u_4')W(y_3|u_4')W(y_4|u_2\oplus u_3'\oplus u_4'),\label{eq:W42}\\
W_4^{(3)}(y_1^4,u_1,u_2|u_3)&=\sum_{u'_4}\frac{1}{8}W(y_1|u_1\oplus u_2\oplus u_4')W(y_2|u_3\oplus u_4')W(y_3|u_4')W(y_4|u_2\oplus u_3\oplus u_4'),\label{eq:W43}\\
W_4^{(4)}(y_1^4,u_1,u_2,u_3|u_4)&=\frac{1}{8}W(y_1|u_1\oplus u_2\oplus u_4)W(y_2|u_3\oplus u_4)W(y_3|u_4)W(y_4|u_2\oplus u_3\oplus u_4),
\label{eq:W44}
\end{align}
where the binary variables $u'_2$, $u'_3$, and $u'_4$ are used to enumerate all possible inputs to the channels.

This channel splitting phase essentially demonstrates how applying one recursion of the generator matrix changes the channel transition probabilities for the channels seen from its input. The generator matrix takes four channels of identical transition probabilities, $W(.|.)$, and produces four new channels with transition probabilities, $W_4^{(i)}(.|.)$, $i=1,2,3,4$. Although the notation is quite straightforward for one recursion, presenting the general recursion expression that takes $N$ channels and produces $4N$ channels is somewhat more complex, as can be observed in \cite{A09}. While the original $2\times 2$ code construction deals with two input sets -- odd and even --  for a channel, the $4\times 4$ case calls for four different sets $G1$, $G2$, $G3$, and $G4$, each representing a set of $N/4$ elements: $G1 = \{1, 5, 9, 13, ...\}$, $G2 = \{2, 6, 10, 14, ...\}$, $G3 = \{3, 7, 11, 15, ...\}$ and $G4 = \{4, 8, 12, 16, ...\}$. Then, the recursive transition probabilities can be obtained as
\begin{align}
W_{4N}^{(4i-3)}(y_1^{4N},u_1^{4i-4}|u_{4i-3})&=\sum_{u_{4i-2}^{4i}}\frac{1}{8}W_N^{(i)}(y_1^N,u_{G1}^{4i-4}\oplus u_{G2}^{4i-4}\oplus u_{G4}^{4i-4}|u_{4i-3}\oplus u_{4i-2}\oplus u_{4i})
\nonumber \\ & \hspace*{30pt} W_N^{(i)}(y_{N+1}^{2N},u_{G3}^{4i-4}\oplus u_{G4}^{4i-4}|u_{4i-1}\oplus u_{4i})
W_N^{(i)}(y_{2N+1}^{3N},
u_{G4}^{4i-4}|u_{4i})\nonumber \\&\hspace*{30pt}W_N^{(i)}(y_{3N+1}^{4N},u_{G2}^{4i-4}\oplus u_{G3}^{4i-4}\oplus u_{G4}^{4i-4}|u_{4i-2}
\oplus u_{4i-1}\oplus u_{4i}),
\label{eq:W4i-3}
\end{align}
\begin{align}
W_{4N}^{(4i-2)}(y_1^{4N},u_1^{4i-4}|u_{4i-3})&=\sum_{u_{4i-1}^{4i}}\frac{1}{8}W_N^{(i)}(y_1^N,u_{G1}^{4i-4}\oplus u_{G2}^{4i-4}\oplus u_{G4}^{4i-4}|u_{4i-3}\oplus u_{4i-2}\oplus u_{4i})\nonumber \\&\hspace*{30pt} W_N^{(i)}(y_{N+1}^{2N},u_{G3}^{4i-4}\oplus u_{G4}^{4i-4}|u_{4i-1}\oplus u_{4i}) W_N^{(i)}(y_{2N+1}^{3N},u_{G4}^{4i-4}|u_{4i})\nonumber \\&\hspace*{30pt}W_N^{(i)}(y_{3N+1}^{4N},u_{G2}^{4i-4}\oplus u_{G3}^{4i-4}\oplus u_{G4}^{4i-4}|u_{4i-2}\oplus u_{4i-1}\oplus u_{4i}),
\label{eq:W4i-2}
\end{align}
\begin{align}
W_{4N}^{(4i-1)}(y_1^{4N},u_1^{4i-4}|u_{4i-3})&=\sum_{u_{4i}}\frac{1}{8}W_N^{(i)}(y_1^N,u_{G1}^{4i-4}\oplus u_{G2}^{4i-4}\oplus u_{G4}^{4i-4}|u_{4i-3}\oplus u_{4i-2}\oplus u_{4i})\nonumber \\&\hspace*{30pt} W_N^{(i)}(y_{N+1}^{2N},u_{G3}^{4i-4}\oplus u_{G4}^{4i-4}|u_{4i-1}\oplus u_{4i}) W_N^{(i)}(y_{2N+1}^{3N},u_{G4}^{4i-4}|u_{4i})\nonumber \\&\hspace*{30pt}W_N^{(i)}(y_{3N+1}^{4N},u_{G2}^{4i-4}\oplus u_{G3}^{4i-4}\oplus u_{G4}^{4i-4}|u_{4i-2}\oplus u_{4i-1}\oplus u_{4i}),
\label{eq:W4i-1}
\end{align}
\begin{align}
W_{4N}^{(4i)}(y_1^{4N},u_1^{4i-4}|u_{4i-3})&=\frac{1}{8}W_N^{(i)}(y_1^N,u_{G1}^{4i-4}\oplus u_{G2}^{4i-4}\oplus u_{G4}^{4i-4}|u_{4i-3}\oplus u_{4i-2}\oplus u_{4i})\nonumber \\&\hspace*{30pt}W_N^{(i)}(y_{N+1}^{2N},u_{G3}^{4i-4}\oplus u_{G4}^{4i-4}|u_{4i-1}\oplus u_{4i})W_N^{(i)}(y_{2N+1}^{3N},u_{G4}^{4i-4}|u_{4i})\nonumber \\&\hspace*{30pt}W_N^{(i)}(y_{3N+1}^{4N},u_{G2}^{4i-4}\oplus u_{G3}^{4i-4}\oplus  u_{G4}^{4i-4}|u_{4i-2}\oplus u_{4i-1}\oplus u_{4i}).
\label{eq:W4i}
\end{align}

\begin{figure}[htbp]
\begin{center}
\includegraphics[width=.8\columnwidth]{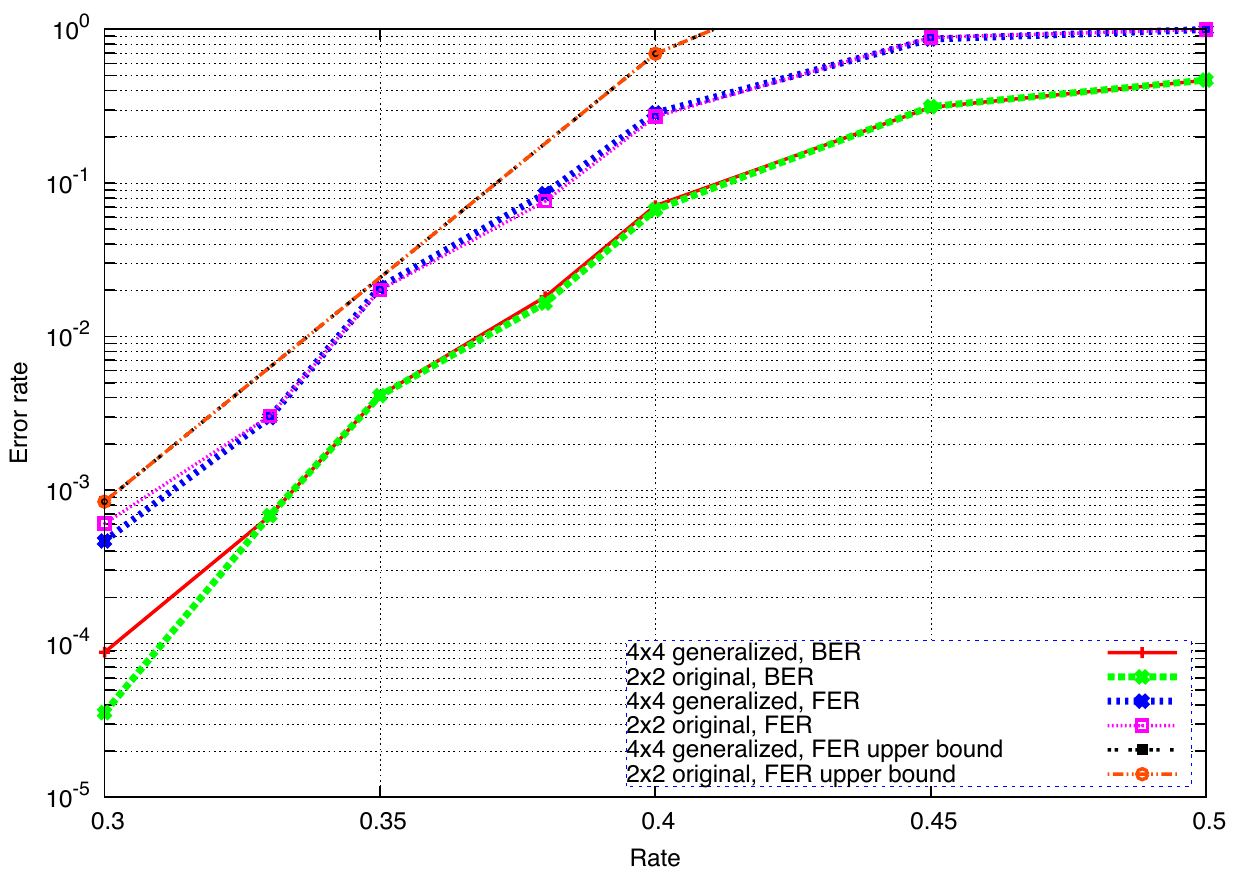}
\end{center}
\caption{BER and FER performance curves for successive cancellation decoding on a BEC with $\epsilon=0.5$.}
\vskip\baselineskip 
\label{fig:decodersim}
\end{figure}

The successive cancellation decoder utilises these channel transition probabilities to estimate the transmitted message. The bit error rate (BER) and frame error rate (FER) performance curves for the successive cancellation decoder for the code designed using $G_e$ are given in Figure \ref{fig:decodersim}. The code lengths for $G_e$ and Arikan's original generator matrix are chosen as 1024, achieved by setting the number of recursions to 5 and 10, respectively. The simulation results verify that the error performance of the codes from Group $1$ is the same as Arikan's original construction's performance. Furthermore, it can be seen that the FER upper bounds for both codes overlap. 

On the implementation side, the decoding graph expands faster for a given fixed number of recursions and this results in a denser graph. In the example code construction we presented above, the desired code length is achievable in just 5 recursions, compared to the 10 recursions of the original code construction. The total number of message updates are the same in both implementations and therefore there are twice the number of message calculations per recursion. This leads to an implementation advantage when parallel processing of multiple messages is possible. In this case, calculating a higher number of messages per recursion would reduce the decoding delay, since the decoder needs to finish one recursion of message calculations before starting the next one.

Another advantage of the proposed code construction lies in the flexibility in the selection of the code length. For practical code lengths to be selected around 1000--8000, a designer utilizing the original code construction method is forced to choose a code length from the set $\{1024, \,2048, \,4096, \,8192\}$ (powers of 2). Using generalised generator matrices, this set expands to
\[ 
\{ 1000, \,1024, \,1296, \,1331, \,1728, \,2048, \,2187, \,2197, \,2401, \,2744, \,3125, \,3375, \,4096, \,6561, \,7776, \,8192\},
\] 
consisting of powers of all small integers. This design flexibility results in an efficient implementation that adjusts to the design requirements of the rest of the communication system.

\section{Conclusion}
\label{conc}

In this work, polarisation performance of polar codes using generalised generator matrices is analysed via both asymptotical and finite-length measures to design and compare efficient polar coded systems. The Bhattacharyya parameters are computed to obtain finite-length measures and the polarisation rate exponents are calculated to obtain asymptotical results over the binary erasure channel. We show that these measures do not always point at the same practical designs and propose a finite-length normalised polarisation distance measure to analytically demonstrate this. Upper bounds on block error probabilities for generator matrices of size  $3\times 3$ and $4\times 4$ are given and it is shown that the bound matches with Arikan's original bound for the best group of size $4\times 4$. The proposed flexible coding structures allow for designing codes of various lengths, not necessarily powers of 2. Although adapting this analysis to other channel models is not straightforward, \cite{A08b} suggests a heuristic method that uses a code designed for the BEC on other channels and reports good experimental results.


\section*{Acknowledgement}
The authors would like to thank the anonymous reviewers for their comments and suggestions that helped improve the presentation of this paper. This work was supported by Bogazici University BAP under grant no 6505.


\begin{thebibliography}{10}

\bibitem{S48}
C.~E. Shannon,
\newblock ``A mathematical theory of communication,''
\newblock {\em Bell Systems Technical Journal}, vol. 27, pp. 379--423, July
  1948.

\bibitem{G62}
R.~G. Gallager,
\newblock ``Low-density parity-check codes,''
\newblock {\em IRE Trans. Inform. Theory}, vol. IT-8, pp. 21--28, Jan. 1962.

\bibitem{G63}
R.~G. Gallager,
\newblock {\em Low-density parity-check codes},
\newblock M.I.T. Press, Cambridge, MA, 1963.

\bibitem{CFRU01}
S.~Y. Chung, G.~D. {Forney, Jr.}, T.~J. Richardson, and R.~L. Urbanke,
\newblock ``On the design of low-density parity-check codes within 0.0045 d{B}
  of the {S}hannon limit,''
\newblock {\em IEEE Communications Letters}, vol. 5, pp. 58--60, Feb. 2001.

\bibitem{A08}
E.~Arikan,
\newblock ``Channel polarization: a method for constructing capacity-achieving
  codes,''
\newblock in {\em Proc. IEEE Intl. Symposium on Inform. Theory}, Toronto,
  Canada, July 2008, pp. 1173--1177.

\bibitem{A09}
E.~Arikan,
\newblock ``Channel polarization: a method for constructing capacity-achieving
  codes for symmetric binary-input memoryless channels,''
\newblock {\em IEEE Trans. Inform. Theory}, vol. IT-55, no. 7, pp. 3051--3073,
  July 2009.

\bibitem{A08b}
E.~Arikan,
\newblock ``A performance comparison of polar codes and reed-muller codes,''
\newblock {\em IEEE Communications Letters}, vol. 12, no. 6, pp. 447--449, June
  2008.

\bibitem{MT09}
R.~Mori and T.~Tanaka,
\newblock ``Performance of polar codes with the construction using density
  evolution,''
\newblock {\em IEEE Communications Letters}, vol. 13, no. 7, pp. 519 --521,
  July 2009.

\bibitem{KSU09}
S.B. Korada, E.~Sasoglu, and R.~Urbanke,
\newblock ``Polar codes: Characterization of exponent, bounds, and
  constructions,''
\newblock in {\em Proc. IEEE Intl. Symposium on Inform. Theory}, Seoul, Korea,
  July 2009, pp. 1483 --1487.

\bibitem{KSU10}
S.B. Korada, E.~{\c S}a{\c s}o{\u g}lu, and R.~Urbanke,
\newblock ``Polar codes: Characterization of exponent, bounds, and
  constructions,''
\newblock {\em IEEE Trans. Inform. Theory}, vol. 56, no. 12, pp. 6253 --6264,
  Dec. 2010.

\bibitem{CNL13}
K. Chen, K. Niu, and J. Lin,
\newblock ``Practical polar code construction over parallel channels,''
\newblock {\em IET Communications}, vol. 7, no. 7, pp. 620 --627,
  May 2013.

\bibitem{AT09}
E.~Arikan and E.~Telatar,
\newblock ``On the rate of channel polarization,''
\newblock in {\em Proc. IEEE Intl. Symposium on Inform. Theory}, Seoul, Korea,
  July 2009, pp. 1493 --1495.

\end{thebibliography}

\end{document}